\documentclass[useams,usenatbib]{mn2e}
\usepackage{graphicx}
\def\lsim{~\raise0.3ex\hbox{$<$}\kern-0.75em{\lower0.65ex\hbox{$\sim$}}~}
\def\gsim{~\raise0.3ex\hbox{$>$}\kern-0.75em{\lower0.65ex\hbox{$\sim$}}~}

\def\lbrack2{[\![}
\def\rbrack2{]\!]}

\def\nx{{N_{x}}}
\def\ny{{N_{y}}}
\def\nz{{N_{z}}}
\def\nangles{{N_{angles}}}

\title[Fully Threaded Transport Engine] {Fully Threaded Transport
Engine: New Method for Multi-Scale Radiative Transfer}

\author[A. O. Razoumov and C. Y. Cardall]
{A. O. Razoumov$^{1,2,3}$\thanks{e-mail: razoumov@phy.ornl.gov} and
C. Y. Cardall$^{1,2}$\thanks{e-mail: ccardall@mail.phy.ornl.gov}\\
$^{1}$ Physics Division, Oak Ridge National Laboratory, Oak Ridge, TN 37831-6354;\\
$^{2}$ Department of Physics and Astronomy, University of Tennessee,
Knoxville, TN 37996-1200\\
$^{3}$ Joint Institute for Heavy Ion Research, Oak Ridge National
Laboratory, Oak Ridge, TN 37831-6374}

\begin{document}

\date{accepted ???. received 2005 March 30; in original form 2005 March 30}

\pagerange{\pageref{firstpage}--\pageref{lastpage}} \pubyear{2002}

\maketitle

\label{firstpage}

\begin{abstract}
A new, very fast method for 3D radiative transfer on fully threaded
grids with arbitrarily high angular resolution is presented. The
method uses completely cell-based discretization, and is ideally
suited for problems with diffuse background radiation, often
encountered in cosmological and star formation models. We find that
for accurate statistical study of intergalactic $Ly\alpha$ absorption
lines one needs of order of few hundred angular discretization
elements even for models without radiative feedback from star forming
galaxies.
\end{abstract}


\section{introduction}

There are many astrophysical systems which require solution of the 3D
radiative transfer equation. In the last decade, starting perhaps from
the local optical depth approximation \citep{gnedin.97}, numerical
transfer solvers started to appear in cosmological and galaxy
formation simulations. A wide range of approximations have been used
to deal with the multi-dimensional nature of the transfer
equation. Among the most interesting original developments one can
mention an implicit variable Eddington tensor method
\citep{norman..98}, a massively parallel multiple wavefront
implementation by the Tsukuba group \citep{umemura.99}, long
characteristics on a uniform grid \citep{razoumov.99}, Monte Carlo
transport \citep{ciardi...01}, optically thin variable Eddington
tensor approach \citep{gnedin.01}, adaptive ray tracing around point
sources \citep{abel.02}, nested trees of rays and sources
\citep{razoumov...02}, and the integral transfer method via fast
Fourier transforms \citep{cen02}. Depending on the problem, many of
these methods effectively reduced the number of operations for
single-frequency transport from $O(N^4 N_{\rm src})$, where $N$ is the
number of data points in each spatial dimension, and $N_{\rm src}$ is
the number of sources, to scaling which is closer to, but not the same
as $O(N^3)$.

Recently, \citet{juvela.05} proposed a new radiative transfer method
for line emission on multi-resolution grids using a combination of
long and short characteristics. In their implementation the radiative
transfer equation is solved on separate grids, and intensities are
interpolated at the grid boundaries. On the other hand,
\citet{cardall...05} developed a method to solve the Boltzmann
equation for neutrino transport on fully threaded grids. The fully
threaded data model pioneered in astrophysical hydrodynamics by
\citet{khokhlov98} is very efficient in dealing with multi-resolution
phenomena. In this paper we extend the ray-tracing scheme of
\citet{juvela.05} to fully threaded structures moving from ray-based
to cell-based discretization, and find that the resulting algorithm is
extremely fast and does not require any significant additional memory
allocation on top of a hydrodynamical solver.

We will show that in general, in order to recover the proper spatial
distribution of the mean intensity in cosmological models, one cannot
(and does not necessarily want to) achieve a scaling much better than
$O(N^3\nangles)$.

We limit transport to plane-parallel fluxes which enter the
computational volume at arbitrary angles, assuming for now that there
are no point sources of radiation in the volume. Extended sources,
i.e. those covering at least few dozen cells can be accurately modeled
with plane-parallel transport with a sufficient number of angles
(current scheme), whereas an unresolved star forming region would need
a separate transfer scheme around point sources on refined meshes
which we have also developed and will present separately. Therefore,
in this paper we limit our tests to a cosmological model with no
feedback.

\section{Method}

\subsection{2D algorithm}

\begin{figure}
  \includegraphics[width=85mm]{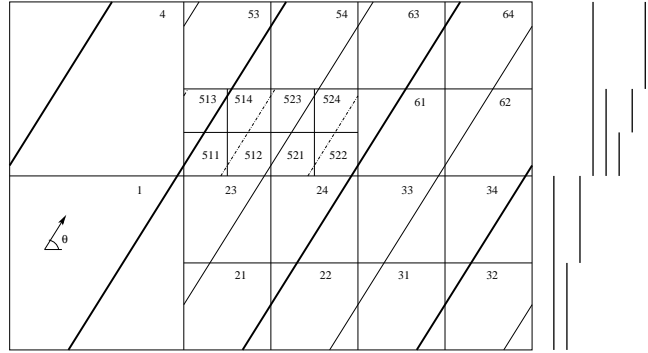} 
  \caption{Ray geometry in 2D with two levels of refinement. The arrow
  shows the photon travel direction. The lines on the right show the
  hierarchy of patterns which hold the ray geometry in the order these
  patterns are created, from left to right. The long rays on the base
  grid are drawn with thick lines, and dash-dotted lines indicate ray
  segments on the second (highest in this case) level of refinement.}
  \label{rayGeometry2D}
\end{figure}

We start with a 2D example to demonstrate the basic ideas of fully
threaded transport, and extend the scheme to 3D in the next
subsection. Consider a 2D Cartesian grid of size $\nx\times\ny$, each
element of which can be refined recursively by a factor of two in each
dimension, resulting in four subcells within a refined element
(Fig.~\ref{rayGeometry2D}). We tag individual cells with a
variable-length label according to their position in the tree
hierarchy.

Assume that a plane-parallel wavefront enters the grid in some
direction $\theta$, which we restrict to $\pi/2<\theta<\pi$. Let us
discretize the wavefront with $\nx$ long characteristics (thick lines
in Fig.~\ref{rayGeometry2D}) separated in $x$ by the base grid
resolution $\Delta x$, so that each characteristic starts in its own
base grid cell at the bottom of the computational grid (cell 1, parent
cell of 21-24, and parent cell of 31-34 in our example), and crosses
the entire grid. Let us now divide each long characteristic into
segments falling into individual cells, and consider those ray
segments to be elements of their parent cells. Each ray segment can be
described by several attributes such as the geometric length and the
optical depth, the orientation of the side it starts on (lower
horizontal or left vertical, for $\pi/2<\theta<\pi$), and the position
of its origin on that side. Using a terminology which will be
particularly useful in 3D, the ray segments starting on the lower
horizontal side of a cell we call the x-rays, and the ones starting on
the left vertical side we call the y-rays. In Fig.~\ref{rayGeometry2D}
the cell 1 has only one x-ray, whereas cell 21 has both an x- and a
y-ray. Note that for our choice of angle $\pi/2<\theta<\pi$ every cell
on any level of refinement will have exactly one x-ray, and zero or
one y-ray. Depending on geometry, these segments might be very short,
e.g., the x-ray in the cell 4, but this ray count rule applies to all
cells on all levels, and is central to our scheme. One can easily
notice that any cell will have only the x-ray and no y-ray if the
x-ray ends at the upper horizontal side of this cell, and an
additional y-ray if the x-ray ends at the right vertical side.

We discretize the rays entirely on the grid storing all information in
computer memory in the form $cell:rayType:rayAttribute$, and each cell
is in turn an element in the grid tree hierarchy. Inside each cell we
measure all lengths and coordinates in units of that cell's size,
independently of its level of refinement.

We now note the second important property of such ray
construction. Since all rays are parallel and separated in $x$ by the
local grid resolution $\Delta x$ (which varies from region to region),
the ray geometry inside each cell -- number of segments, their lengths
and orientation -- is the same for all cells of a given refinement
level within each horizontal layer. For example, in
Fig.~\ref{rayGeometry2D} cells 21, 22, 31 and 32 have the same ray
geometry, and so do the cells 511, 512, 521, 522. Similarly, the cells
513, 514, 523 and 524 have their own geometry. It immediately follows
that the ray geometry has to be computed only once per level, per
layer.

We start calculation by computing the ray geometry of cell 1 and
recording it in pattern $P_1$. Next we sub-divide this pattern into
two elements $P_{11}$ and $P_{12}$, the first one containing the ray
geometry of cells 21, 22, 31 and 32, and the second one -- ray
geometry of cells 23, 24, 33 and 34. In the second base grid layer we
have one additional level of refinement, and store the patterns $P_2$,
$P_{21}$, $P_{211}$, $P_{212}$, and $P_{22}$. In each cell we record a
pointer to its pattern, although this association can also be
recovered from position of that cell in the tree, with a few extra
integer operations. In addition, we compute the optical depth along
each ray element.

Before actual transport, we introduce the concept of x- and
y-neighbours of each cell which are the cells into which the x- and
y-ray ``back-trace'' into. Neighbours can be of the higher, lower, or
same level of refinement as the current cell. Cell 1 in
Fig.~\ref{rayGeometry2D} does not have any neighbours since its only
x-ray starts on the grid boundary. Cell 21 has 1 as its y-neighbour,
since its y-ray starts on the interface with cell 1, but it has no
x-neighbour. Cell 511 has one x-neighbour (23) and one y-neighbour
(4). Cell 53 has one x-neighbour (514) and one y-neighbour (4), and so
on. For each cell $C$ we store its x- and y-neighbours as pointers
$C:{\cal X}$ and $C:{\cal Y}$.

Next we perform transport, starting from cell 1, continuing in cells
21, 22, 23, 24, 31, 32, 33, 34, 4, 511, 512, and so on until we cover
the entire grid. In each cell we start by computing transfer of the
intensity along the x-ray. The input intensity $C:xray:I_{\rm in}$ is
either the cosmic background intensity (cells 1, 21, 22, 31, 32,
i.e. those along the lower edge of the grid), or the outgoing
intensity of the respective x-neighbour, either $C:{\cal X}:xray$ or
$C:{\cal X}:yray$, depending on which ray segment hits the horizontal
interface between $C$ and $C:{\cal X}$. Note that by construction only
one ray segment in $C:{\cal X}$ reaches this interface, so there is no
ambiguity in the choice of the incoming intensity $C:xray:I_{\rm
in}$. When we continue a ray from a lower-resolution into a
higher-resolution cell, i.e.  when $C:level > C:{\cal X}:level$, we do
not interpolate across the wavefront and instead just propagate the
``un-refined intensity'' $C:{\cal X}:xray:I_{\rm out}$. We found that
in practice the error associated with this approximation is
negligible, since it is in the \emph{refined region} that radiation
starts to show a non-zero gradient across the wavefront on scales
smaller than the size of $C:{\cal X}$.

Going back to our example, to get the input x-ray intensity in both
the cells 511 and 512, we use the outgoing x-ray intensity from their
x-neighbour, cell 23. In cell 53 the incoming x-ray intensity is the
outgoing y-ray intensity from cell 514 (the thick line in
Fig.~\ref{rayGeometry2D}). After an x-ray update in each cell we
proceed to a y-ray update in that cell if a y-ray segment is
present. The incoming y-ray intensity is again either the cosmic
intensity for border cells (cell 4), or the outgoing x-ray (only!)
intensity of the corresponding y-neighbour. Note that by construction
a y-ray in any cell never reaches the cell's right vertical side,
always ending at the upper horizontal side. The only special case for
a y-update which does not arise in x-updates is when $C:{\cal Y}:xray$
ends on the upper horizontal side of $C:{\cal Y}$, instead of the
right vertical side. This in fact happens in cell 21 for which its
y-neighbour's x-ray (cell 1) does not end on the right vertical
side. In this case we use $C:yray:I_{\rm in}=cell:{\cal Y}:xray:I_{\rm
out}$.

\subsection{3D unigrid implementation}

\begin{figure}
  \includegraphics[width=65mm]{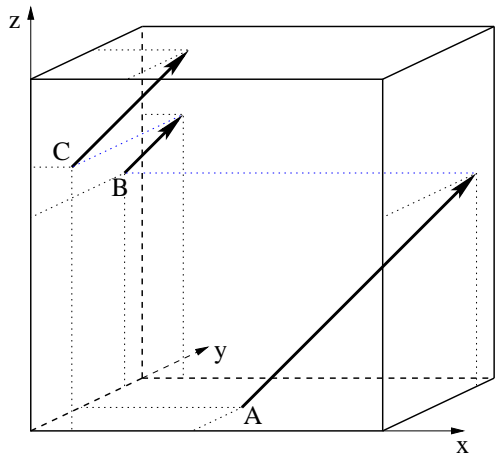} 
  \caption{Ray geometry in a 3D cell. There is always one xy-ray which
  starts at point A (lower xy-face), and for this particular orientation the
  cell also contains a yz-ray starting at point B (left yz-face), and an
  xz-ray starting at point C (front xz-face).}
  \label{rayGeometry3D}
\end{figure}

Now consider a 3D uniform Cartesian grid of size
$\nx\times\ny\times\nz$ and a plane-parallel wavefront entering the
volume in some direction $\phi,\theta$. As in the 2D case, we limit
the angles $\phi,\theta$ in such a way that a ray entering a cubic
cell through its lower left front corner ($x=y=z=0$ in
Fig.~\ref{rayGeometry3D}) would leave this cell through its upper
xy-face without hitting the side faces first. Such restriction on the
angles means that for now $\phi,\theta$ can cover only a solid angle
of $\pi/6$. We discretize the wavefront with $\nx\times\ny$ long
characteristics, each one of which starts in its own base grid cell at
the bottom ($z=0$) of the computational volume and crosses the entire
grid until it leaves the volume. In analogy with the 2D scheme each
long characteristic is divided into cell segments, but now each
segment can start on a xy-plane, xz-plane, or yz-plane
(Fig.~\ref{rayGeometry3D}). We continue the notation from the previous
section and call these segments the xy-, xz-, and yz-rays,
respectively, and consider them to be elements of their parent
cells. As before, $C_{i,j,k}:xyray:x_0$ stands for the x-coordinate of
the starting point of the xy-ray inside cell $C_{i,j,k}$, and we
assume that inside each cell all lengths are measured in units of that
cell's size, so that the starting position of each long characteristic
is described by the same pair of numbers $C_{1,j,k}:xyray:x_0,y_0$
which do not depend on indices $j,k$, and we wrote $i=1$ since all
long characteristics start in the bottom layer cells.

As long as $\phi,\theta$ stay inside the solid angle we chose above,
each base grid cell in the volume will have exactly one xy-ray in it,
zero or one xz-ray, and zero or one yz-ray. Since the grid is
Cartesian, and we have fixed the angles, all base grid cells in each
layer $i$ have exactly the same set of rays segments, which can be
conveniently stored in a pattern $P_i$, to which each cell in that
layer has a pointer $C_{i,j,k}:P\rightarrow P_i$. We have to compute
and store this pattern only once per layer. The starting points
$P_{i+1}:xyray:x_0,y_0$ for each subsequent layer can be computed
entirely from $P_i$.

Without refinement one can compute radiative transfer one layer at a
time, starting from the very bottom layer $i=1$. In each layer, except
for $i=1$, the first step would be to compute a pair of starting
points $P_i:xyray:x_0,y_0$ from $P_{i-1}$, and then calculate the
entire pattern $P_i$ starting from the length of the xy-ray and the
type of the plane it finishes on (xy, xz or yz). If the xy-ray ends on
the xy-plane, then it is the only ray segment in this
pattern. However, if it finishes on the xz-plane, there will be an
xz-ray inside this pattern (Fig.~\ref{rayGeometry3D}). If the xz-ray
is present, it will end on either the xy- or yz-plane, but not on
xz-plane, due to the restriction in angles we chose above. In the
latter case we'll have one yz-ray in this pattern, and so on. For each
additional ray segment we need to compute its starting point, either
$P_i:xzray:x_0,z_0$ or $P_i:yzray:y_0,z_0$, and the length of that ray
segment. We also record, for quick reference, the type of the ray that
hits each of the planes (xy, xz, yz) in that cell.

Once the layer pattern is computed, we can use it for every cell in
that layer. In each cell $C_{i,j,k}$ we start by doing transfer along
the xy-ray. The starting intensity is either the cosmic background
intensity ($i=1$), or the outgoing intensity of the ray segment (xy,
xz, or yz) in $C_{i-1,j,k}$ that hits the upper xy-plane in that
cell. An angle-dependent source function discretized on this ray
segment can also be added to the transport step. Next, if the segment
$C_{i,j,k}:xzray$ is active, we perform corresponding transfer of the
incoming intensity, which is taken to be either the cosmic background
($j=1$), or the outgoing intensity of the ray segment in $C_{i,j-1,k}$
that ends on the xz-plane of that cell. A similar update is done for
the segment $C_{i,j,k}:yzray$ if it exists in this cell.

\subsection{3D implementation with refinement}

Let us finally consider a fully threaded 3D data structure in which
each cell $C_{i,j,k}$ can be refined, in which case it contains eight
subcells $C_{i,j,k}:C_{i',j',k'}$ of size 1/2 of the parent, and
assume that this refinement can proceed to any arbitrary level. Then
the ray pattern $P_i$ for each layer will also be recursively refined
to form structures of the type $P_i:P_{i'}$, up to the deepest level
of cell refinement in that layer. Unlike in the unrefined case, each
ray segment is not necessarily a simple continuation of corresponding
ray segment in a neighbouring cell. If we just entered a refined
region, new ray segments are created, and there are several strategies
one can use to refine intensities across the wavefront. In complete
analogy with the 2D construction above, we choose not to interpolate
the intensities, but just to use the outgoing intensity of the cell
into which the refined ray ``back-traces''. However, we need an extra
step to find out which cells the xy-, xz- and yz-rays ``back-trace''
into, or what we call the xy-, xz- and yz-neighbours of the current
cell. For each ray segment present, we store the pointers to their
respective neighbours, which can be of the same, higher or lower level
of refinement. Note that every cell containing, e.g., an xz-ray must
have an xz-neighbour, unless it starts on the boundary of the
computational volume.

As in the unrefined scheme, before we perform transfer in a particular
layer, we need to compute the base grid ray pattern for that layer. We
then proceed to perform transfer in all base grid cells, following
xy-rays and, if present, xz-rays and yz-rays, from their respective
neighbours. When we hit a refined cell, we need to compute the
corresponding refined ray pattern to which this cell has a pointer, if
that pattern has not been computed before, and do that for each cell
and refinement level in the layer. Each pattern at each level of
refinement needs to be computed only once per layer.

The radiative transfer calculation itself is very fast, since all we
need to do is follow the interconnected data structures of cells and
ray segments we have already created for a particular choice of angles
$\phi,\theta$. We always compute radiative transfer on the deepest
level of refinement available, thus eliminating the need to store
intensities at the boundary between regions of different refinement
levels. The only floating point operations we need are the pattern
calculation ($O(\nz)$), the calculation of the optical depth $\tau$ in
each cell ($O(\nx\times\ny\times\nz)$), and the optical depth
attenuation $e^{-\tau}$ ($O(\nx\times\ny\times\nz)$). If we include
emission, we also need to compute the source function
($O(\nx\times\ny\times\nz)$). There are one, two or three ray segments
in each cell, the average number falling between one and two and
depending on the grid size and hierarchy complexity. The total number
of operations per iteration or timestep is then
$O(\nx\times\ny\times\nz\times\nangles)$, plus some overhead for
refinement.

To generalize this scheme for any direction $\phi,\theta$, we just
need to rotate the indices used to access the grid elements, both on
the base grid and in refined cells, to cover all 24 angular zones of
$\pi/6$, three (xy, xz, or yz) for each of the eight octants in a
cube.

For discretization in angles we use the HEALPix algorithm
\citep{gorski...02}, dividing the entire sphere into $12\times4^{n-1}$
equal area pixels, where $n=1,2,...$, so that in computing the mean,
angle-averaged intensity (which then goes into the ionizational
balance module) all quadrature terms have the same weight. The
angle-dependent intensity in each cell is obtained by averaging over
all ray segments in a given direction present in that cell. We do not
correct for the fact that the center of the ray segment does not
coincide with the center of the cell, since in our discretization the
intensity is viewed more as a cell-averaged quantity. In addition, we
find that averaging over few (one, two or three) ray segments cancels
the errors faster than interpolation to the cell center.

Finally, the directional intensity along each ray segment is computed
as the average over that segment

\begin{equation}
\bigl<I\bigr>_{\rm seg}={I_{\rm in}\over L}\int_0^L e^{-\kappa l}dl=
{I_{\rm in}\over\kappa L}\left(1-e^{-\kappa L}\right)=
{I_{\rm in}-I_{\rm out}\over\ln(I_{\rm in}/I_{\rm out})},
\end{equation}

\noindent
where $L$ is the geometrical length, and $I_{\rm in}$ and $I_{\rm
out}$ are the incoming and attenuated intensities at the ends of that
segment. Using this form instead of the simple average $(I_{\rm
in}+I_{\rm out})/2$ ensures that we get a correct answer in cells
which have large absorption but are nevertheless exposed to a strong
radiation background.

\section{Application}

\begin{figure*}
  \includegraphics[scale=1.0]{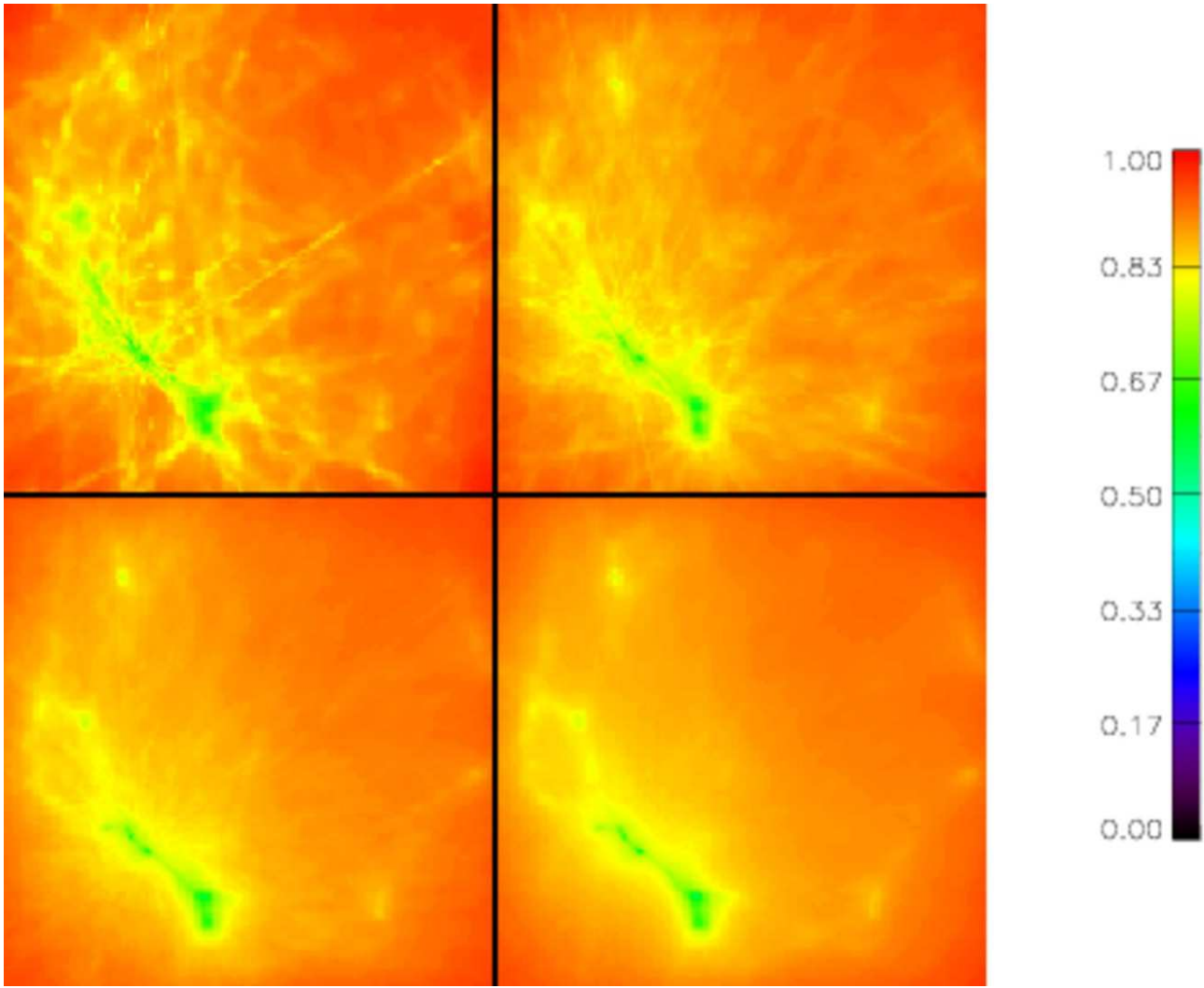} 
  \caption{Color map of the mean intensity in a slice through the
    8 Mpc (comoving) volume, in units of the mean background
    intensity, at $z=3$, for a $128^3$ run with 5 levels of refinement
    by a factor of two. The upper left panel shows the calculation
    with 12 angles, the upper right -- with 48 angles, and the bottom
    panels are with 192 and 768 angles, respectively (see text for
    details). Only 3 levels of refinement are plotted for
    simplicity.}
  \label{intensity2e28}
\end{figure*}

We ran a number of standard tests of the type described in
\citet{razoumov.99} to make sure that we recover the proper mean
intensity on uniform grids. To test the method with cell refinement,
we took an output from a cosmological simulation computed with
\emph{Enzo}, an Eulerian AMR structure formation code
\citep{bryan.99}, in the 8 Mpc (comoving) volume at $z=3$ at $128^3$
base grid resolution, assigned some constant cross-section $\sigma$
which is typical -- within a factor of two -- for photons above the
Lyman limit propagating in the intergalactic medium, and ran a simple
optical attenuation calculation with the variable number of angles
$12\times4^{n-1}=12,48,192$ and 768. The entire dataset included 501
nested grids with up to five levels of refinement, which for the
purposes of our transfer we projected onto a fully threaded data
structure. Full calculation -- computing ray geometry, finding
neighbours and doing actual transport -- takes about 3 seconds in one
direction on the full tree hierarchy on a 2 Ghz Pentium 4
processor. In Fig.~\ref{intensity2e28} we plotted the angle-averaged
intensity in a cross-section through one of the dense halos, in units
of the mean cosmic background.

At lower angular resolution the radiation field is extremely clumpy
forming long shadows behind shielded and semi-shielded
regions. Visually, the quality of the solution improves drastically
when we switch from 48 to 192 angles, and the latter seems to be an
almost converged solution as there is very little difference in
radiation fields computed at 192 and 768 angles. We conclude that for
a volume of several Mpc on a side at $z=3$ one needs at least a
hundred angular resolution elements to compute the diffuse
component. In general, this number depends on the size of the volume,
the redshift and the photon frequency.

\begin{figure}
  \includegraphics[width=80mm]{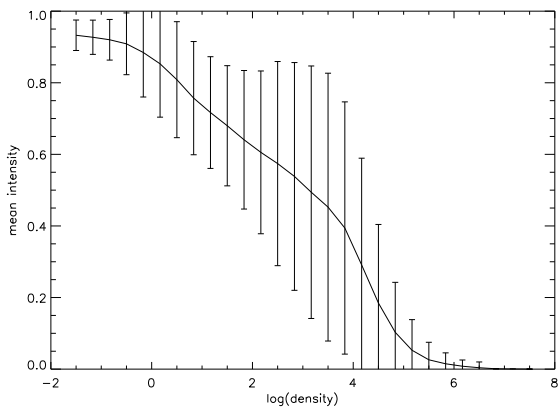} 
  \caption{Mean intensity, in units of the mean background intensity, as
  a function of density, in units of the mean density in the box,
  grouped into 30 bins. The errorbars show r.m.s. dispersion in each bin.}
  \label{intensityDensity}
\end{figure}

Fig.~\ref{intensityDensity} shows the mean intensity as a function of
the local density using data from all cells on all levels of
refinement, for a high angular resolution model. The vertical
errorbars demonstrate a large r.m.s. mean intensity dispersion for all
cells of the given density, indicating that radiative transfer effects
are not negligible in modeling the thermal state of the intergalactic
medium.

\section{Discussion}

In our implementation, since transfer is done layer by layer, there is
no need to store any transfer-related variables in more than two base
grid layers at any given time, the current layer and the one below. Of
course, one would prefer to keep grid variables such as density,
temperature and ionization state of the entire volume in memory at
once, otherwise these variables would need to be read from the disk
every new timestep or iteration. Overall, if all cells are kept in
memory, our implementation does not require any more storage than a
hydrodynamical part. For parallel implementation, if the entire cell
hierarchy can fit on one processor, parallelization via angle
decomposition is preferable to volume decomposition.

In a hydrodynamical structure formation code, radiative transfer is
followed by calculation of ionizational balance and gas
temperature. We find that for large hydrodynamical time steps 10-20
iterations are normally required to compute an equilibrium position of
HI, HeI and HeII ionization fronts. Interestingly enough, fewer
iterations are needed to obtain a solution in higher spatial
resolution models in which cosmological filaments and sheets are
better resolved and tend to leave more open space for radiation to
reach the higher density regions.

\section*{Acknowledgments}

A.R. would like to thank \AA ke Nordlund for organizing the radiative
transfer and MHD workshop in Copenhagen in December 2004 which
inspired this work, and Mika Juvela and Eric Lentz for stimulating
discussions. This work was supported by Scientific Discovery Through
Advanced Computing (SciDAC), a program of the Office of Science of the
U.S.  Department of Energy (DoE); and by Oak Ridge National
Laboratory, managed by UT-Battelle, LLC, for the DoE under contract
DE-AC05-00OR22725.

\bibliography{ref}

\begin{thebibliography}{}

\bibitem[\protect\citeauthoryear{{Abel} \& {Wandelt}}{{Abel} \&
  {Wandelt}}{2002}]{abel.02}
{Abel} T.,  {Wandelt} B.~D.,  2002, MNRAS, 330, L53

\bibitem[\protect\citeauthoryear{{Bryan} \& {Norman}}{{Bryan} \&
  {Norman}}{1999}]{bryan.99}
{Bryan} G.~L.,  {Norman} M.~L.,  1999, in Baden S.~B.,  Chrisochoides N.~P.,
  Gannon D.,   Norman M.~L.,  eds, IMA Vol. 117, Structured Adaptive Mesh
  Refinement (SAMR) Grid Methods, New York: Springer, p.~165

\bibitem[\protect\citeauthoryear{{Cardall}, {Razoumov}, {Endeve} \&
  {Mezzacappa}}{{Cardall} et~al.}{2005}]{cardall...05}
{Cardall} C.~Y.,  {Razoumov} A.~O.,  {Endeve} E.,    {Mezzacappa} A.,  2005, in
  Mezzacappa A.,  Fuller G.~M.,  eds, {Open Issues in Core Collapse Supernova
  Theory,} Singapore: World Scientific

\bibitem[\protect\citeauthoryear{{Cen}}{{Cen}}{2002}]{cen02}
{Cen} R.,  2002, ApJS, 141, 211

\bibitem[\protect\citeauthoryear{{Ciardi}, {Ferrara}, {Marri} \&
  {Raimondo}}{{Ciardi} et~al.}{2001}]{ciardi...01}
{Ciardi} B.,  {Ferrara} A.,  {Marri} S.,    {Raimondo} G.,  2001, MNRAS, 324,
  381

\bibitem[\protect\citeauthoryear{{G{\' o}rski}, {Banday}, {Hivon} \&
  {Wandelt}}{{G{\' o}rski} et~al.}{2002}]{gorski...02}
{G{\' o}rski} K.~M.,  {Banday} A.~J.,  {Hivon} E.,    {Wandelt} B.~D.,  2002,
  in Bohlender D.~A.,  Durand D.,   Handley T.~H.,  eds, Astronomical Data
  Analysis Software and Systems XI, ASP Conference Series, Vol. 281, p.~107

\bibitem[\protect\citeauthoryear{{Gnedin} \& {Abel}}{{Gnedin} \&
  {Abel}}{2001}]{gnedin.01}
{Gnedin} N.~Y.,  {Abel} T.,  2001, New Astronomy, 6, 437

\bibitem[\protect\citeauthoryear{{Gnedin} \& {Ostriker}}{{Gnedin} \&
  {Ostriker}}{1997}]{gnedin.97}
{Gnedin} N.~Y.,  {Ostriker} J.~P.,  1997, ApJ, 486, 581

\bibitem[\protect\citeauthoryear{{Juvela} \& {Padoan}}{{Juvela} \&
  {Padoan}}{2005}]{juvela.05}
{Juvela} M.,  {Padoan} P.,  2005, ApJ, 618, 744

\bibitem[\protect\citeauthoryear{{Khokhlov}}{{Khokhlov}}{1998}]{khokhlov98}
{Khokhlov} A.~M.,  1998, J. Comput. Phys., 143, 519

\bibitem[\protect\citeauthoryear{{Norman}, {Paschos} \& {Abel}}{{Norman}
  et~al.}{1998}]{norman..98}
{Norman} M.~L.,  {Paschos} P.,    {Abel} T.,  1998, Memorie della Societa
  Astronomica Italiana, 69, 455

\bibitem[\protect\citeauthoryear{{Razoumov}, {Norman}, {Abel} \&
  {Scott}}{{Razoumov} et~al.}{2002}]{razoumov...02}
{Razoumov} A.~O.,  {Norman} M.~L.,  {Abel} T.,    {Scott} D.,  2002, ApJ, 572,
  695

\bibitem[\protect\citeauthoryear{{Razoumov} \& {Scott}}{{Razoumov} \&
  {Scott}}{1999}]{razoumov.99}
{Razoumov} A.~O.,  {Scott} D.,  1999, MNRAS, 309, 287

\bibitem[\protect\citeauthoryear{{Umemura}, {Nakamoto} \& {Susa}}{{Umemura}
  et~al.}{1999}]{umemura.99}
{Umemura} M.,  {Nakamoto} T.,    {Susa} H.,  1999, in Miyama S.~M.,  Tomisaka
  K.,   Hanawa T.,  eds, ASSL Vol. 240: Numerical Astrophysics, {3D Radiative
  Transfer Calculations of the Cosmic Reionization}.
Boston, Mass. : Kluwer Academic, p.~43

\end{thebibliography}

\bsp

\label{lastpage}

\end{document}